# A Search for Strongly Mg-enhanced Stars from the Sloan Digital Sky Survey

Xiang Li[1,2], Gang Zhao[1], Yu-Qin Chen[1], Hai-Ning Li[1]

[1] Key Laboratory of Optical Astronomy, National Astronomical Observatories, Chinese Academy of Sciences, Beijing 100012, China; *gzhao@nao.cas.cn*

[2] University of Chinese Academy of Sciences, Beijing 100049, China

**Abstract** Strongly Mg-enhanced stars with [Mg/Fe] > 1.0 show peculiar abundance patterns and hence are of great interest for our understanding of stellar formation and chemical evolution of the Galaxy. A systematical search for strongly Mg-enhanced stars based on the low-resolution ($R \simeq 2000$) spectra of the Sloan Digital Sky Survey (SDSS) is carried out by finding the best matched synthetic spectrum to the observed one in the region of Mg I b lines around $\lambda$5170Å via a profile matching method. The advantage of our method is that fitting parameters are refined by reproducing the [Mg/Fe] ratios of 47 stars from very precise high-resolution spectroscopic (HRS) analysis by Nissen & Schuster (2010); and these parameters are crucial to the precision and validity of the derived Mg abundances. As a further check of our method, Mg abundances are estimated with our method for member stars in four Galactic globular clusters (M92, M13, M3, M71) which cover the same metallicity range as our sample, and the results are in good agreement with those of HRS analysis in the literature. The validation of our method is also proved by the agreement of [Mg/Fe] between our values and those of HRS analysis by Aoki et al. (2013). Finally, 33 candidates of strongly Mg-enhanced stars with [Mg/Fe]>1.0 are selected from 14850 F and G stars. Follow-up observations will be carried out on these candidates with high-resolution spectroscopy by large telescopes in the near future, so as to check our selection procedure and to perform a precise and detailed abundance analysis and to explore the origins of these stars.

## 1 INTRODUCTION

Enhancements of $\alpha$-elements in metal-poor stars have been first identified by Aller & Greenstein (1960), and later Wallerstein (1962) has confirmed this enhancement as a typical [$\alpha$/Fe] ratio of $\sim +0.40$ at [Fe/H]$\sim -1.0$ based on a sample of G type dwarf stars in the disk. Subsequently, more investigations have shown that metal-poor halo stars almost have a constant [α/Fe] ratio of ∼ +0.4 (McWilliam 1997; Zhao et al. 1998; Zhao & Gehren 2000; Gehren et al. 2004; Zhang et al. 2009). It is also noticed that [α/Fe] ratios of thick-disk stars( ∼ +0.3 to +0.4) are higher than those of thin-disk stars (∼ +0.1 to +0.2) (Bensby et al. 2005; Reddy et al. 2006). However, some strongly Mg-enhanced stars with [Mg/Fe] ratio larger than 1.0 have been found among extremely metal-poor (EMP) stars (Aoki et al. 2000, 2002a,b, 2005, 2006, 2007a,b, 2013; Cohen et al. 2006, 2011, 2013). In particular, a well-studied star, CS 22949-037 with [Mg/Fe]= +1.26 is found to share similar abundance as normal metal-poor stars for other α-elements including Ca and Ti, as well as light odd-Z elements such as Na and Al, but exhibit peculiar abundance of neutron-capture elements. It seems that the nucleosynthesis mechanisms for these strongly Mg-enhanced stars are quite different from those of normal stars.

It has been suggested that there are three categories of strongly Mg-enhanced stars. (1) Carbon-enhanced metal-poor stars with large overabundant elements produced by the slow- and rapid-neutron-capture process (CEMP-rs). $^{25}$Mg and $^{26}$Mg are produced significantly in high-mass AGB stars during a convective s-process driven by the $^{22}$Ne(α,n)$^{25}$Mg neutron source (Goriely & Siess 2005; Karakas & Lattanzio 2003), which is responsible for the observed high Mg (=$^{24}$Mg+$^{25}$Mg+$^{26}$Mg) abundance in some CEMP-rs stars (Masseron et al. 2010). (2) Carbon-enhanced metal-poor stars with no enhancement of neutron-capture elements (CEMP-no). CEMP-no stars are born out of gas with large amount of C, which is polluted by low-energy faint supernova, and they have undergone the first dredge-up and processed certain amount of pristine C into N (Ryan et al. 2005). The high [Mg/Fe] ratio of CEMP-no stars can be explained with the "mixing and fallback" model by Umeda & Nomoto (2005), which suggests that the high [Mg/Fe] ratio occurs if the mixing-fallback region (with large amount of ejected Fe) does not extend beyond the Mg layers (Tsujimoto & Shigeyama 2003). (3) α-enhanced metal-poor star with no enhancement of carbon and neutron-capture elements (AEMP). The explanation for the only AEMP star, BS 16934-002, is analogous to that of CEMP-no stars, except that the non-enhancement of C is due to the effect of significant mass loss of out layers containing C-rich material produced from its massive progenitor (Aoki et al. 2007b). The origins of the strongly Mg-enhanced stars and their abundance patterns provide us crucial evidence to understand the early chemical evolution of the Galaxy.

However, the number of strongly Mg-enhanced stars identified by now is still very limited and there is not yet any systematic search for such stars. The large database from the SDSS spectroscopic survey (York et al. 2000) provides us an unprecedented opportunity to conduct such a systematic investigation of strongly Mg-enhanced stars; therefore our work aims to utilize the spectroscopic data from SDSS DR9 to conduct a systematic search for strongly Mg-enhanced stars (Ahn et al. 2012). This paper is organized as follows. In section 2, we give a brief description on the sample selection. Section 3 describes the method for [Mg/Fe] determinations. A list of the candidates of strongly Mg-enhanced stars is presented in Section 4, and Section 5 is a summary of the main results.

## 2 THE SAMPLE SELECTION

The spectroscopic data and atmospheric parameters we have used are based on SDSS DR9. Although the latest data release of DR10 (Ahn et al. 2014) is available now, there is no update on the low-resolution stellar spectroscopy and thus it would make no difference if the sample is re-selected from DR10. The selection procedure is as follows. Firstly, F and G-type stars with $0.2 < (g-r)_0 < 0.55$ (Yanny et al. 2009) and S/N at g-band larger than 30 are selected. Secondly, based on the stellar parameters determined by SSPP (SEGUE Stellar Parameter Pipeline, Lee et al. 2008a,b; Allende Prieto et al. 2008), further selection are made to include only stars with [Fe/H] $< -1.0$ $dex$ and 5000 K $<$ T$_{eff}$ $<$ 7000 K. It is because strongly Mg-enhanced stars reported in literatures are all metal-poor stars and SSPP tends to provide more reliable stellar parameters in this range of temperature. According to Ahn et al. (2012), for a typical G-type dwarf in the color range of $0.4 < g-r < 1.3$ with S/N per pixel of 30, the internal uncertainties of SSPP parameters are $\sim$50 K for T$_{eff}$, $\sim$0.12 $dex$ for log $g$, and $\sim$0.1 $dex$ for [Fe/H], respectively. These errors increase to $\sim$80 K, $\sim$0.3 $dex$, and $\sim$0.25 $dex$ for T$_{eff}$, log $g$ and [Fe/H] for stars with $-0.3 < g-r < 0.2$, [Fe/H] $< -2.0$, and S/N $<$ 15. Finally, we estimate the S/N around Mgb lines, S/N$_{Mg}$ from SDSS spectra in the wavelength range of 5280−5325 Å, because some stars with S/N $>$ 30 (given by SSPP) are not with sufficient spectral quality at Mg I b region which is important for our work. Thus, we exclude stars with S/N$_{Mg}$ $<$ 30. The above selection results in a final sample of 14,850 stars.

## 3 THE DETERMINATION OF MAGNESIUM-TO-IRON RATIOS

Spectral synthesis of the Mg I b feature is carried out to derive the [Mg/Fe] ratios of the sample with the local thermodynamic equilibrium (LTE) atmospheric models. Although Lee et al. (2011) have determined [$\alpha$/Fe] ratios from the SDSS spectra of these stars, values of individual stars are not publically available at present. Moreover, their [$\alpha$/Fe] ratios are derived from four $\alpha$-elements, Mg, Ti, Si and Ca over the wide wavelength range from 4500-5500 Å by using the weighting factors 5, 3, 1 and 1, respectively. The general matching to a wide wavelength range of spectra may be a good approximation for estimating [$\alpha$/Fe], but has no advantage in searching for strongly Mg-enhanced candidates since the contributions from other elements in the wide wavelength range will possibly counterbalance the contribution of the enhancement of Mg elements (if exists) to an undetected level. This is also pointed out by Lee et al. (2011) that such measurement may not correctly represent the overall content of the $\alpha$-elements, especially in cases of abnormally high or low Mg abundances.

In this paper, we aim to derive [Mg/Fe] ratios from the narrow wavelength range around the Mg I b feature, which is dominated by Mg element, with a specific purpose of picking out strongly Mg-enhanced candidates. A line-profile-matching method is performed on the three Mg lines individually by varying [Mg/Fe] ratios of the synthetic spectra from $-0.6$ to $+2.0$. The $\chi^2$ values of a set of [Mg/Fe] ratios are fitted with a third order polynomial fit and the [Mg/Fe] corresponding with the minimum $\chi^2$ value is considered to be the best-fit value. The matching procedure with $\chi^2$ method is widely used, but the choice of different fitting parameters is key to the derived abundances. In the following sections, we will focus on our efforts in refining these fitting parameters by using the spectra and the results from high-resolution analysis.

### 3.1 Observational and Synthetic Spectra

In order to set the fitting parameters and to check whether the Mg I b feature is a valid and robust proxy for estimating [Mg/Fe] ratio, we apply our matching method to a reference sample of 47 dwarf stars (Nissen &

Schuster 2010), (hereafter NS10), with high-resolution and high-S/N FIES (FIbre fed Echelle Spectrograph) spectra (kindly provided by Poul Nissen) together with accurate atmospheric parameters and very high precision Mg abundances. The FIES spectra of 47 reference stars cover a wavelength range from 4000 Å to 7000 Å with a resolution of $R \simeq 40000$ and $S/N \simeq 140-200$.

The synthetic spectra are generated by a reliable and user-friendly code SPECTRUM (v2.76) written by Richard O. Gray[1]. The code adopts the one-dimensional, 72-layer, plane-parallel and line-blanketed models without convective overshoot, which are linearly interpolated over $\alpha$-enhanced AODFNEW grid (Castelli & Kurucz 2003). The [$\alpha$/Fe] is assumed to be $0.4\ dex$ when [Fe/H] $< -0.5\ dex$ in these models.

The detailed procedure of our method is as follows. According to atmospheric parameters provided by NS10, corresponding atmosphere model is generated by linear interpolations with Castelli and Kurucz's model grid. After that, a set of synthetic spectra are produced by SPECTRUM for various [Mg/Fe] ratios from $-0.6$ to $+2.0$. Then the high resolution synthetic spectra with $R \simeq 50000$ are reduced to the resolution similar to the observed spectra by an auxiliary program of SPECTRUM named CUSTOMSM. The $\chi^2$ values are calculated from the deviation between the synthetic and the observed spectra. A third order polynomial is used to fit $\chi^2$ values, and the [Mg/Fe] ratio corresponding to the minimum $\chi^2$ value is considered to be the best-fit value.

### 3.2 Set the fitting parameters

Before applying our method to the SDSS spectra, we would like to refine and testify our method with high-resolution and high-S/N FIES spectra. Firstly, the line list and atomic data of the Mg I b region should be checked. An isotope-compatible line list provided by SPECTRUM is adopted as it is updated recently (private communication with Richard O. Gray). However, we find that the $log\ gf$ value of Fe 5162.292 Å is not suitable, which will blend the first line of Mg I b in low-resolution spectra. Because the synthetic spectra with accurate Fe abundances from NS10 do not match this iron line for any of the 47 FIES spectra. We thus adopt the $log\ gf$ value of this line from Lambert et al. (1996), which results in a good agreement between the observed and synthetic spectra. Secondly, we check the reliability of [Mg/Fe] derived from Mg I b compared with other two magnesium lines, Mg 4730.04 Å and Mg 5711.10 Å which are often used in abundance analysis of high-resolution spectra. We apply our line-profile-matching procedure to Mg 4730.04 Å, Mg 5711.10 Å and Mg I b lines of FIES spectra with $R \simeq 40000$, and obtain [Mg/Fe] ratios very close to the results of NS10 with a deviation less than $0.1\ dex$. The [Mg/Fe] derived from these three lines are presented in columns 7 to 9 of Table 1, and the deviation between the [Mg/Fe] determined from Mg I b and the other two magnesium lines is about $0.05\ dex$. According to Gehren et al. (2006), such deviation can be explained by the NLTE effect which is around $0.065\ dex$. Thirdly, we apply our method to the smoothed FIES spectra with $R \simeq 2000$ which is the same as SDSS spectra to check the validity of our method on low-resolution spectra. The smooth factor should be fixed as an input parameter for SPECTRUM when we degrade the initial synthetic spectra with $R \simeq 50000$ to $R \simeq 2000$. To do this, we select three high S/N, SDSS-I calibration stars which have well determined stellar parameters and $\alpha$ abundances based on high-resolution spectra observed with the Hobby Eberly Telescopy (Lee et al. 2011). We adjust the smooth factor until the smoothed synthetic spectrum well matches the observed one and a final value of 2.2 is adopted as the optimum value as shown in Figure 1. The [Mg/Fe] derived from the smoothed FIES spectra with $R \simeq 2000$ is presented in column 10 of Table 1 and the values are very close to the results using Mg I b of high-resolution FIES spectra. Thus we can apply our method to the low-resolution SDSS spectra by using the Mg I b lines to obtain [Mg/Fe] ratio. Figure 2 shows the match of the three magnesium lines between the synthetic spectra and the observed one for G176-53 in NS10.

---

[1] http://www.appstate.edu/ grayro/spectrum/spectrum.html

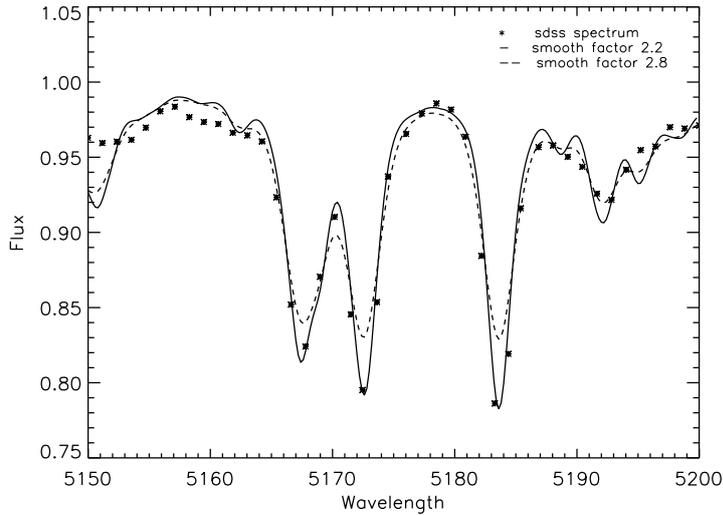

**Fig. 1** Comparison of a SDSS spectrum and corresponding synthetic spectra with different smooth factors (solid line for 2.2 and dash line for 2.8)

Finally, the possible effect on the derived [Mg/Fe] from the uncertainties of micro-turbulence is considered. We change the micro-turbulence value of NS10 by $\pm 0.3$ kms$^{-1}$ and derive [Mg/Fe] ratios with the smoothed FIES spectra. It is found that such variations in the micro-turbulence will not significantly affect the [Mg/Fe] derived from the smoothed FIES spectra of NS10. Figure 3 presents the difference of [Mg/Fe] rations derived from the low-resolution spectra due to a change of $\pm 0.3$ kms$^{-1}$ in micro-turbulence for the 47 FIES stars in NS10. The resulted deviation in [Mg/Fe] is less than 0.03 $dex$.

### 3.3 Calibration with Globular Cluster

Some Galactic globular cluster with precise abundance analysis of high-resolution have been observed by SDSS. Therefore, it will be a good test to apply our method on the SDSS spectra of member stars of globular cluster. Member stars from four Galactic globular clusters (M92, M13, M3, M71) with a coverage in the metallicity space comparable to our sample are carefully chosen to test the validation of our method. Member stars in clusters are supposed to be born simultaneously and exhibit similar elemental-abundance patterns due to the well-mixed interstellar medium at the same location in the Galaxy. Although the anti-correlations of Na-O and Mg-Al in globular clusters (GCs) indicate significant star-to-star variations of these light elements, according to Carretta et al. (2009), the variation of Mg abundances in 19 normal GCs is around 0.07 $dex$ for HB stars. The four GCs adopted here have been well investigated in the literature and the scatters in their [Mg/Fe] ratios are around $0.1-0.2$ $dex$ as shown in Table 2, which may be mainly caused by the uncertainty of the analysis and thus may not be their intrinsic scatters. Since the scatters in the four GCs are not significant and comparable with the uncertainty of our [Mg/Fe] determination, the derived [Mg/Fe] ratios for member stars of these GCs can testify the application of our method on SDSS data.

**Table 1** Comparison of [Mg/Fe] for different Mg lines and different resolutions of Mg I b feature for stars in NS10

| ID | $T_{eff}$ | log $g$ | [Fe/H] | Turb | [Mg/Fe]$_{NS10}$ | HRS-4703 | HRS-5711 | HRS-Mg I b | LRS-Mg I b |
|---|---|---|---|---|---|---|---|---|---|
| G05-36 | 6013 | 4.23 | -1.233 | 1.39 | 0.33 | 0.34 | 0.33 | 0.24 | 0.24 |
| G119-64 | 6181 | 4.18 | -1.477 | 1.50 | 0.25 | 0.23 | 0.20 | 0.10 | 0.12 |
| G125-13 | 5848 | 4.28 | -1.434 | 1.50 | 0.30 | 0.30 | 0.20 | 0.20 | 0.21 |
| G127-26 | 5791 | 4.14 | -0.529 | 1.21 | 0.31 | 0.21 | 0.34 | 0.20 | 0.19 |
| G13-38 | 5263 | 4.54 | -0.876 | 0.92 | 0.36 | 0.34 | 0.45 | 0.35 | 0.35 |
| G15-23 | 5297 | 4.57 | -1.097 | 1.00 | 0.40 | 0.34 | 0.44 | 0.39 | 0.39 |
| G150-40 | 5968 | 4.09 | -0.807 | 1.41 | 0.14 | 0.18 | 0.18 | 0.09 | 0.08 |
| G16-20 | 5625 | 3.64 | -1.416 | 1.51 | 0.22 | 0.25 | 0.25 | 0.20 | 0.21 |
| G161-73 | 5986 | 4.00 | -0.999 | 1.36 | 0.13 | 0.04 | 0.10 | 0.00 | 0.00 |
| G170-56 | 5994 | 4.12 | -0.924 | 1.49 | 0.18 | 0.14 | 0.17 | 0.05 | 0.06 |
| G172-61 | 5225 | 4.47 | -1.000 | 0.86 | 0.16 | 0.22 | 0.24 | 0.25 | 0.25 |
| G176-53 | 5523 | 4.48 | -1.337 | 1.00 | 0.15 | 0.21 | 0.21 | 0.18 | 0.19 |
| G180-24 | 6004 | 4.21 | -1.393 | 1.55 | 0.31 | 0.29 | 0.29 | 0.23 | 0.24 |
| G187-18 | 5607 | 4.39 | -0.666 | 1.15 | 0.29 | 0.27 | 0.35 | 0.26 | 0.26 |
| G192-43 | 6170 | 4.29 | -1.339 | 1.45 | 0.19 | 0.20 | 0.17 | 0.09 | 0.09 |
| G20-15 | 6027 | 4.32 | -1.485 | 1.60 | 0.22 | 0.21 | 0.27 | 0.16 | 0.15 |
| G21-22 | 5901 | 4.24 | -1.089 | 1.40 | 0.09 | 0.06 | 0.06 | 0.02 | 0.01 |
| G232-18 | 5559 | 4.48 | -0.928 | 1.26 | 0.36 | 0.34 | 0.40 | 0.31 | 0.31 |
| G24-13 | 5673 | 4.31 | -0.721 | 0.96 | 0.34 | 0.24 | 0.40 | 0.26 | 0.26 |
| G24-25 | 5825 | 3.85 | -1.402 | 1.13 | 0.35 | 0.37 | 0.39 | 0.21 | 0.21 |
| G31-55 | 5638 | 4.30 | -1.097 | 1.36 | 0.28 | 0.34 | 0.37 | 0.32 | 0.31 |
| G49-19 | 5772 | 4.25 | -0.552 | 1.22 | 0.30 | 0.24 | 0.35 | 0.23 | 0.22 |
| G53-41 | 5859 | 4.27 | -1.198 | 1.30 | 0.24 | 0.28 | 0.22 | 0.15 | 0.14 |
| G56-30 | 5830 | 4.26 | -0.891 | 1.32 | 0.09 | 0.07 | 0.14 | 0.05 | 0.04 |
| G56-36 | 5933 | 4.28 | -0.938 | 1.43 | 0.20 | 0.17 | 0.24 | 0.12 | 0.13 |
| G57-07 | 5676 | 4.25 | -0.474 | 1.09 | 0.34 | 0.25 | 0.40 | 0.27 | 0.26 |
| G74-32 | 5772 | 4.36 | -0.724 | 1.14 | 0.37 | 0.27 | 0.39 | 0.26 | 0.25 |
| G75-31 | 6010 | 4.02 | -1.035 | 1.38 | 0.12 | 0.14 | 0.10 | 0.05 | 0.04 |
| G81-02 | 5859 | 4.19 | -0.689 | 1.31 | 0.25 | 0.15 | 0.26 | 0.11 | 0.11 |
| G85-13 | 5628 | 4.38 | -0.586 | 0.97 | 0.33 | 0.27 | 0.38 | 0.26 | 0.25 |
| G87-13 | 6085 | 4.13 | -1.088 | 1.52 | 0.10 | 0.16 | 0.08 | 0.05 | 0.06 |
| G94-49 | 5373 | 4.50 | -0.796 | 1.10 | 0.35 | 0.33 | 0.39 | 0.36 | 0.35 |
| G96-20 | 6293 | 4.41 | -0.889 | 1.52 | 0.30 | 0.26 | 0.24 | 0.17 | 0.17 |
| G98-53 | 5848 | 4.23 | -0.874 | 1.30 | 0.20 | 0.19 | 0.24 | 0.17 | 0.15 |
| G99-21 | 5487 | 4.39 | -0.668 | 0.89 | 0.33 | 0.28 | 0.42 | 0.28 | 0.27 |
| HD148816 | 5823 | 4.13 | -0.731 | 1.43 | 0.32 | 0.27 | 0.37 | 0.23 | 0.23 |
| HD159482 | 5737 | 4.31 | -0.726 | 1.31 | 0.34 | 0.29 | 0.38 | 0.26 | 0.25 |
| HD160693 | 5714 | 4.27 | -0.487 | 1.12 | 0.24 | 0.20 | 0.34 | 0.18 | 0.17 |
| HD177095 | 5349 | 4.39 | -0.737 | 0.85 | 0.38 | 0.27 | 0.46 | 0.30 | 0.30 |
| HD179626 | 5850 | 4.13 | -1.041 | 1.57 | 0.35 | 0.34 | 0.41 | 0.25 | 0.27 |
| HD189558 | 5617 | 3.80 | -1.121 | 1.39 | 0.36 | 0.32 | 0.40 | 0.27 | 0.27 |
| HD193901 | 5650 | 4.36 | -1.090 | 1.22 | 0.13 | 0.13 | 0.18 | 0.11 | 0.11 |
| HD194598 | 5942 | 4.33 | -1.093 | 1.40 | 0.18 | 0.16 | 0.22 | 0.09 | 0.10 |
| HD230409 | 5318 | 4.54 | -0.849 | 1.11 | 0.30 | 0.29 | 0.38 | 0.33 | 0.33 |
| HD233511 | 6006 | 4.23 | -1.547 | 1.30 | 0.36 | 0.32 | 0.27 | 0.29 | 0.29 |
| HD237822 | 5603 | 4.33 | -0.450 | 1.09 | 0.35 | 0.27 | 0.41 | 0.28 | 0.27 |
| HD250792 | 5489 | 4.47 | -1.013 | 1.08 | 0.23 | 0.20 | 0.25 | 0.20 | 0.24 |

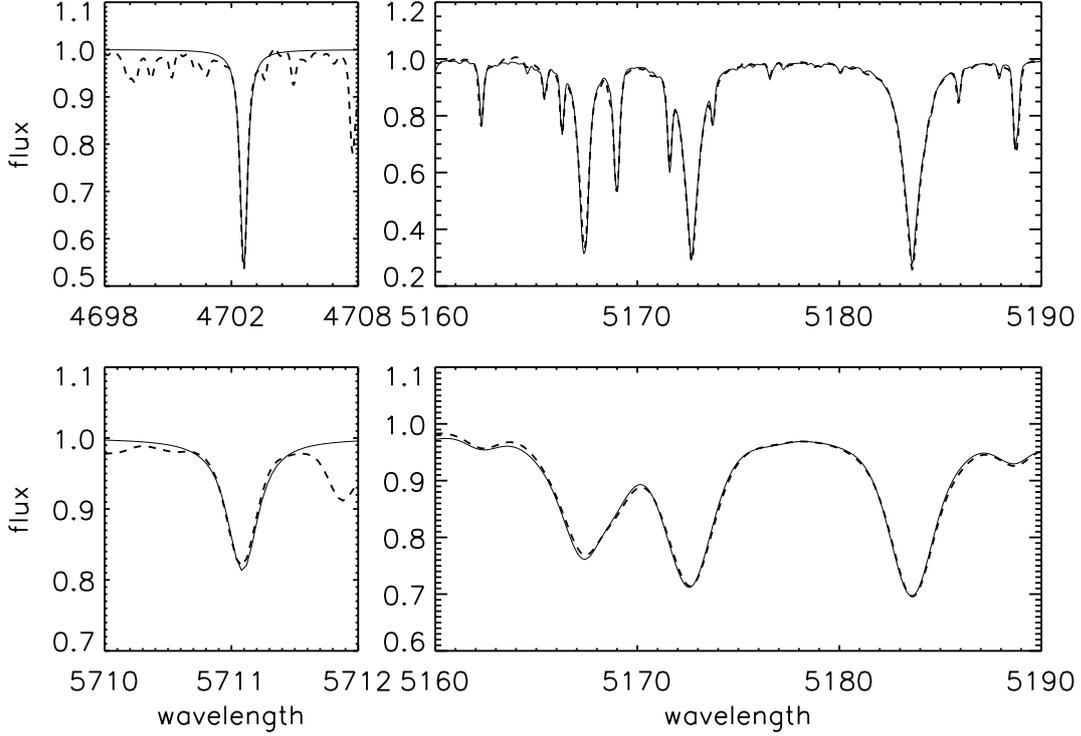

**Fig. 2** Upper and lower left panels: The matching of Mg 4703 and Mg 5711 lines between FIES spectrum (solid lines) and the best matched synthetic spectrum (dash lines) for G176-53. Upper right panels: The matching of Mg I b lines between the FIES spectrum (solid lines) and the best matched synthetic spectrum (dash lines). Lower right panel: The matching of Mg I b lines between the degraded FIES spectrum (solid lines) and the best matched synthetic spectrum (dash lines).

The member lists of these GCs from SDSS were given by Smolinski et al. (2011), and we select member stars with high quality spectra of $S/N_{Mg} > 30$. For these stars, micro-turbulences are not available, and we adopt Equation 1 for $\log g > 4.0$ based on Bruntt et al. (2010) and Equation 2 for $\log g < 4.0$ which is obtained by performing a fitting function to $T_{eff}$ and $\xi_t$ data using UVES giant and sub-giant samples (Lind et al. 2009). The uncertainty of the micro-turbulence is about 0.3 kms$^{-1}$, which only has a negligible effect on the derived [Mg/Fe] ratio as discussed in Sect. 3.2 and also shown in Figure 3. We apply the line-profile-matching method to the member stars of the four GCs, and obtain the mean [Mg/Fe] and its scatter for each cluster by averaging individual values of member stars. The detailed procedure for the determination of [Mg/Fe] from the SDSS spectra will be described in Sect. 4.

$$\xi_t = 1.01 + 4.56 \times 10^{-4}(T_{eff} - 5700) + 2.75 \times 10^{-7}(T_{eff} - 5700)^2 \qquad (1)$$

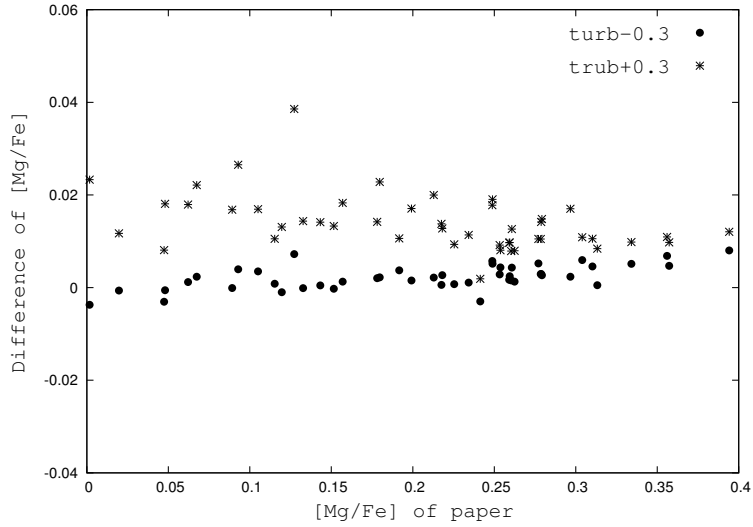

**Fig. 3** The deviations of [Mg/Fe] ratios by varying $\xi_t$ with $\pm 0.3$ kms$^{-1}$ for 47 stars in NS10.

**Table 2** Stellar parameters and [Mg/Fe] ratios of four globular clusters by HRS analysis

| Cluster | Num(stars) | [Fe/H] | $\sigma_{[Fe/H]}$ | [Mg/Fe] | $\sigma_{[Mg/Fe]}$ | Reference[a] |
|---|---|---|---|---|---|---|
| M92 | 6 | -2.31 | 0.08 | +0.19 | 0.19 | S96 |
| M13 | 25 | -1.50 | 0.05 | +0.24 | 0.15 | C05 |
| M3 | 13 | -1.39 | 0.05 | +0.40 | 0.12 | C05 |
| M71 | 24 | -0.78 | 0.10 | +0.36 | 0.09 | RC02 |

Notes: [a] S96:Shetrone 1996; C05:Cohen & Meléndez 2005; RC02:Ramírez & Cohen 2002

$$\xi_t = 8.02 - 2.17 \times 10^{-3} T_{eff} + 1.74 \times 10^{-7} (T_{eff})^2 \qquad (2)$$

Stellar parameters and mean [Mg/Fe] ratios of the four clusters are adopted from the literatures which are based on high-resolution abundance analysis as listed in Table 2. The comparisons of [Mg/Fe] from our work and HRS studies are shown in Figure 4. Although the scatter in [Mg/Fe] of the four GCs is about 0.12 $dex$, there is a systematic shift to a higher value than previous work by about 0.13 $dex$. We suspect that the difference between atmospheric parameters derived from SSPP and those by HRS studies could explain the observed systematic shift. Figure 5 shows the distribution of the T$_{eff}$ versus log $g$ of the GCs' member stars compared with Dartmouth isochrones[2] with metallicities and age close to those of each GC. It is shown that for M13, M3 and M71, the HRS sample nicely follow the isochrones, while our samples show a noticeable deviation, with +0.5 $dex$ in log $g$ and $-250$ K in T$_{eff}$. We thus re-calculate the [Mg/Fe] ratios of sample stars of these GCs by varying log $g$ by $+0.5$ $dex$ and T$_{eff}$ by $-250$ K, and find a difference in [Mg/Fe] of 0.26 $dex$ and 0.2 $dex$, respectively, which is able to explain the shift in [Mg/Fe]. We also take this effect into account in the following selection of strongly Mg-enhanced stars.

---

[2] http://stellar.dartmouth.edu/models/index.html

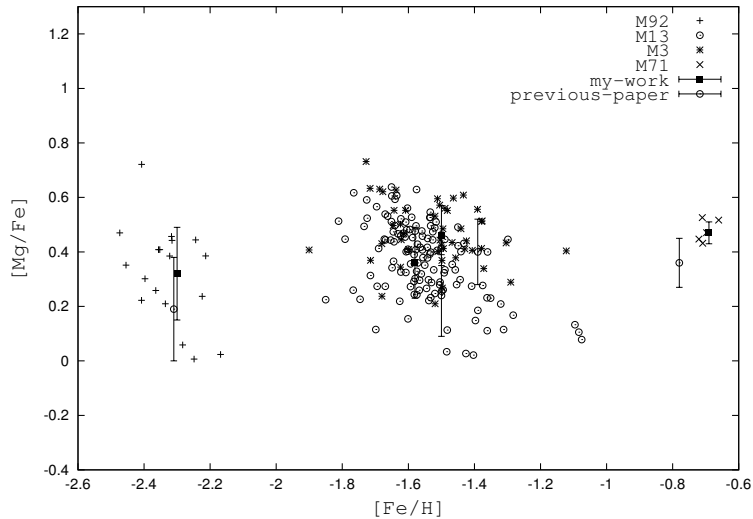

**Fig. 4** The comparison of [Mg/Fe] of four globular clusters between our work and HRS analysis. The mean [Mg/Fe] and rms are 0.47±0.04, 0.46±0.11, 0.36±0.13, and 0.32±0.17 for M71, M3, M13, and M92, respectively.

### 3.4 Test with High Resolution Spectral Analysis

In order to make a direct comparison, we apply our method on SDSS spectra for 136 extremely metal-poor stars which has been observed by the High Dispersion Spectrograph of the Subaru Telescope and analysed in details by Aoki et al. (2013) (hereafter Aoki13). Our derived [Mg/Fe] ratios are then compared with those from Aoki13. Among the 136 stars, 45 stars are excluded due to low S/N$_{Mg}$ ($< 30$) and another 12 stars are excluded due to their weak Mg I b feature or bad continuum from the blended C$_2$ band. A final sample for comparison includes 80 stars, resulting in an average deviation of 0.21 $dex$ between our work and Aoki13 with a scatter of 0.24 $dex$. We check the distribution of these 80 stars on the diagram of T$_{eff}$ versus log $g$, and find 19 stars located outside one sigma region of the theoretical isochrone, which are considered with unreliable atmosphere parameters and then excluded from the comparison sample. The remaining 61 stars with S/N$_{Mg}$ >30 and reliable atmospheric parameters can be used to make a star-to-star comparison of [Mg/Fe]. The left panel of Figure 6 shows a good agreement between the two sets of data. There is no obvious deviation from the one-to-one line, and the scatter of 0.18 $dex$ is within the typical error of our measurements. Note that for the 61 stars, the atmospheric parameters derived form SSPP (which we have adopted) are not exactly the same as those by Aoki13, but there should not be large differences after checking the locations of these objects in the T$_{eff}$ versus log $g$ diagram. But the spectra of the two works and the procedure of deriving [Mg/Fe] are very different: our values are derived from low-resolution spectra via the line-profile-matching method, while Aoki13 uses high-resolution spectra and measures the [Mg/Fe] ratio from individual magnesium lines. The consistency in the [Mg/Fe] ratios between our work and Aoki13 indicates that our method of deriving [Mg/Fe] from low-resolution SDSS spectra is reliable. In particular, stars with high [Mg/Fe]($> 0.8$ $dex$) in Aoki13 are well reproduced by our method.

Out of the 136 stars, there are 122 with [$\alpha$/Fe] ratios available from the SDSS DR7 database. The right panel of Figure 6 compares the [$\alpha$/Fe] derived from SSPP DR7 and [Mg/Fe] from Aoki13. Apparently, the agreement is not so good and there is no object with [$\alpha$/Fe] larger than 0.4 dex if the SSPP [$\alpha$/Fe] is adopted. So indeed, described in Lee et al. (2011), the measured [$\alpha$/Fe] from averaging four individual $\alpha$-elements by using weighting factors may not correctly represent the overall content of specific element like Mg. Therefore, the [$\alpha$/Fe] from SSPP cannot be used to search for strongly Mg-enhanced stars.

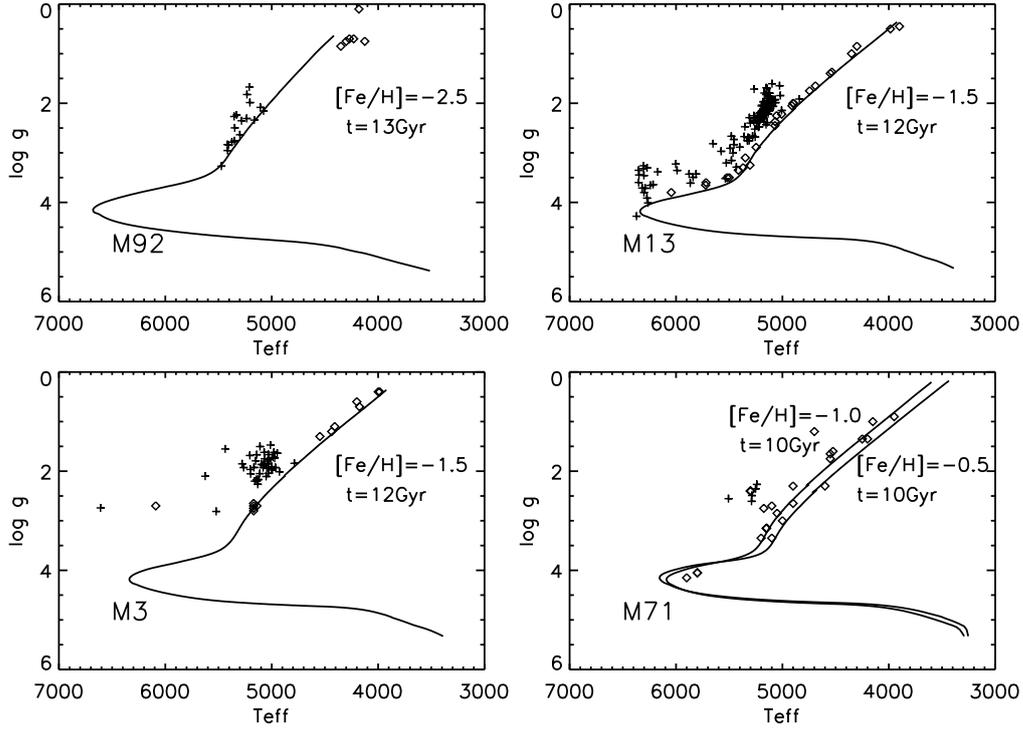

**Fig. 5** The log $g$ versus $T_{eff}$ diagram of member stars of four GCs with the updated Dartmouth isochrones (2012 version) from Dotter et al. (2008) (solid lines). Stars from HRS analysis are shown in rhombuses and SDSS cluster members in pluses.

### 3.5 Internal uncertainty

The uncertainties of the measured [Mg/Fe] come from the spectra, the models and the errors of atmospheric parameters, which can be quantitatively estimated from a Monte Carlo simulation. We choose two stars, SDSS J134922.91+140736.9 and SDSS J130538.1+194305.6, as examples of our sample. In particular, one is on the turnoff stage and the other is on the red giant branch. Moreover, SDSS J134922+140736 is confirmed as a strongly Mg-enhanced star (Sbordone et al. 2012). Then we perform the Monte Carlo simulation for the two stars based on 500 sets of atmospheric parameters, which are generated with the errors of atmospheric parameters of SSPP with $\pm 150\,K$, $\pm 0.3\,dex$ and $\pm 0.3\,dex$ for $T_{eff}$, log $g$ and [Fe/H], respectively. Tests are made and proved that 500 sets of atmospheric parameters are enough for this simulation while larger sets of data will not result much differently. The mean value and rms of the simulation is $+1.16 \pm 0.28\,dex$ for SDSS J134922.91+140736.9, and $+0.97 \pm 0.30\,dex$ for SDSS J130538.1+194305.6, as shown in Figure 7. It is worth noticing that, the small scatter of 0.18 $dex$ on [Mg/Fe] between our result and Aoki et al. (2013) shows that the external error may not be as large as 0.2 $dex$, and the internal error of 0.3 $dex$ estimated by the Monte Carlo simulation should be the upper limit of the uncertainty.

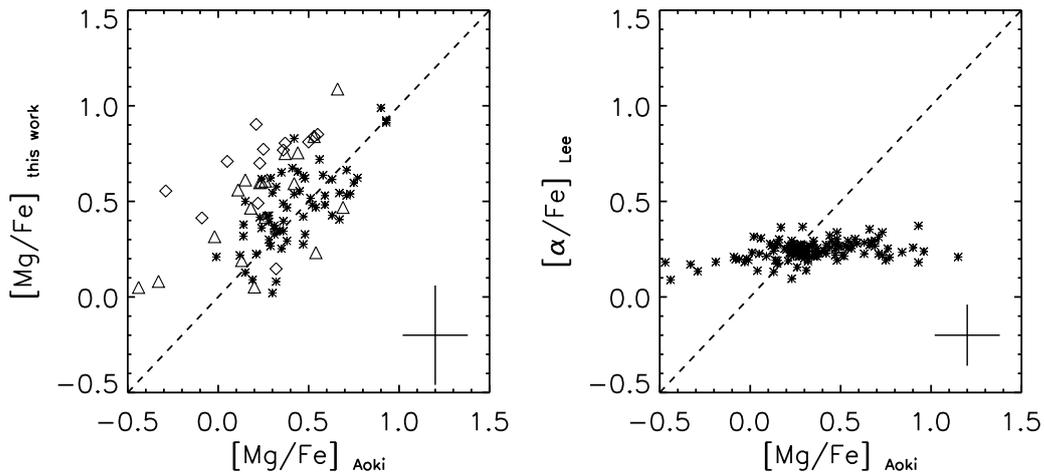

**Fig. 6** The comparisons of [Mg/Fe] from our work (left panel) and Lee et al. (2011) (right panel) with the HRS from Aoki et al. (2013). Left panel: 12 stars (rhombuses) are excluded due to their weak Mg I b feature or bad continuum from the blended $C_2$ band of SDSS spectra, and 19 stars (triangles) are excluded due to their unreliable atmospheric parameters from SSPP.

## 4 THE SELECTION OF STRONGLY MG-ENHANCED CANDIDATES

The determination of [Mg/Fe] ratios from SDSS spectra is carried out by the following two steps. Firstly, we transform vacuum-based SDSS spectra into the air-based scale and shift the spectra to the rest frame with the radial velocity of SSPP. Secondly, we select 12 continuum windows in the wavelength range of 4900 Å - 5400 Å based on a set of high-resolution spectra provided by the ELODIE archive with stellar atmospheric parameters covering the similar range to that of our sample. We obtain the continuum by a polynomial fit to these windows of the whole wavelength range of 4900 Å - 5400 Å , and normalize the SDSS spectra by dividing the spectra with the continuum. Finally, the synthetic spectra are normalized in the same way to ensure the observed SDSS spectra and the synthetic spectra can be well matched. Following the method and fixed parameters described in Sect. 3.2, the [Mg/Fe] ratios are derived for our 14,850 sample stars and their distributions of [Mg/Fe] versus [Fe/H] are shown in Figure 8.

Among our sample, 174 stars are with [Mg/Fe] greater than 1.0 and located above the dotted line in Figure 8. Individual spectra of the 174 stars are checked by eyes and stars with unclear Mg I b feature or not-well-defined continuum are excluded. Finally, 84 strongly Mg-enhanced candidates are selected through the interactive checking shown with asterisks in Figure 8. In particular, SDSS J134922+140736 ,the strongly Mg-enhanced star discovered by Sbordone et al. (2012) is included in our candidate list; however, SDSS J084016+540526 discovered by Aoki et al. (2013) is not selected into our sample due to its low S/N$_{Mg}$ (<30) in the SDSS spectra. Note that the high [Mg/Fe] of these candidates may be incorrectly estimated if their atmospheric parameters from SSPP are unreliable. Therefore, in order to check their atmospheric parameters ,we compare the 84 candidates in the T$_{eff}$ versus log $g$ diagram with isochrones from Dotter et al. (2008) with corresponding metallicities in the possible age range of 10-14 Gyr. According to their locations in Figure 9, 33 candidates with reliable atmospheric parameters from SSPP are finally picked out and are listed in Table 3. These 33 candidates are further divided into two types: type A with atmospheric parameters fairly following the isochrones and type B with the location falling within one sigma error region. These final selected 33 candidates are shown with red squares in Figure 8. Considering that the deviation of atmospheric parameters from isochornes would produce an uncertainty of [Mg/Fe] of 0.3 $dex$, while our selection criterion of [Mg/Fe]>1.0 is three times larger than the normal [Mg/Fe] ratio of +0.3 ∼ +0.4 $dex$ for most EMP stars, we could still expect type B to be qualified candidates of strongly Mg-enhanced stars.

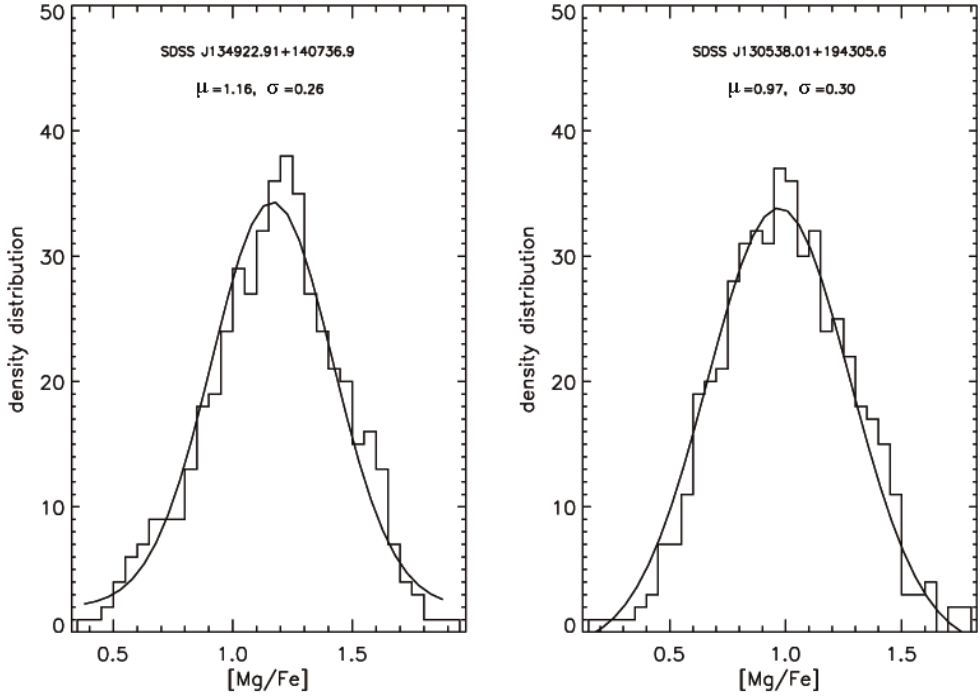

**Fig. 7** The Monte Carlo simulation of the [Mg/Fe] distribution for SDSS J134922+140736 ($T_{eff}$=6342 K, log $g$=3.89, [Fe/H]=−2.83, [Mg/Fe] =+1.16±0.26), and SDSS J130538+194305 ($T_{eff}$=5230.05 K, log $g$=2.46, [Fe/H]=−2.25, [Mg/Fe] = +0.97±0.30).

It is known that the strength of Mg I b lines are sensitive to the log $g$ and if the log $g$ is underestimated, the Mg abundance will be overestimated. Besides, most of members of the GCs used for calibration in section 3.3 are giants and subgiants, while most of our 33 candidates are turn-off and dwarf stars. Therefore, we have further checked the effect of the assumed underestimation of log $g$ on the derived [Mg/Fe] ratios of the final 33 candidates. According to Figure 5, it seems SSPP underestimated log $g$ by 0.5 $dex$ which is added to the log $g$ of the 33 candidates. As shown in Figure 10, the effects of increase of log $g$ on [Mg/Fe] of turn-off and dwarf stars are slightly larger than those of giants and sub-giants in the order of 0.03 $dex$. For the turn-off and dwarf stars, the average decrease of the [Mg/Fe] to the assumed increase of log $g$ is 0.28 $dex$. If such an decrease in [Mg/Fe] is taken into account, the fractions of [Mg/Fe] greater than 1.0, 0.9, 0.8 and 0.7 of our 33 candidates are 34%, 49%, 93% and 100%, respectively. Even under the extreme case of underestimation of log $g$ by 0.5 $dex$, the [Mg/Fe] of all the 33 candidates are still obviously larger than those of normal stars and hence our selection is confirmed to be reliable.

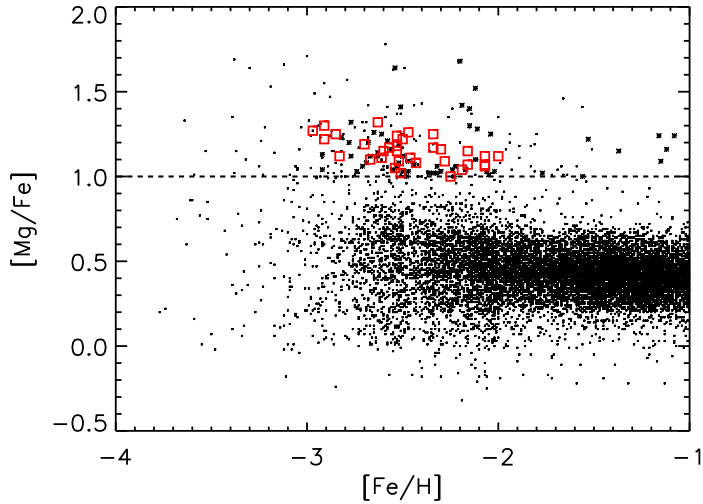

**Fig. 8** The [Fe/H] versus [Mg/Fe] diagram of the sample stars. Asterisks are 84 strongly Mg-enhanced candidates after interactive checking of clear Mg I b line profile and well-defined continuum. The red squares are the 33 strongly Mg-enhanced candidates after atmospheric parameters check.

## 5 SUMMARY

Based on a line-profile-matching method applied on the Mg I b feature of SDSS spectra for a sample of 14,850 F and G stars, 33 strongly Mg-enhanced candidates with [Mg/Fe]>1.0 are discovered. This is the first systematic search for strongly Mg-enhanced candidates. With the quality of spectra and the uncertainty of stellar parameters checked, as well as the measurement error estimated from comparison with high-resolution spectral analysis and Monte Carlo simulation, the selected candidates are proved to be reliable.

In particular, there are obvious advantages of the line-profile-matching method in searching for strongly Mg-enhanced candidates. First, we use high-resolution spectra from NS10 to check the line list and to set the fitting parameters as well as estimation of dependence of the adopted micro-turbulence on the [Mg/Fe] determined from the Mg I b feature. The difference between our [Mg/Fe] ratios derived from the Mg I b feature and NS10's values derived from other weak Mg lines is less than 0.1 $dex$, indicating the validation of our method. Moreover, we derive [Mg/Fe] from the Mg I b feature for member stars of four globular cluster observed in SDSS and they are consistent with those from previous HRS investigations. The method presented here can be used to the spectral analysis of the LAMOST (The Large Sky Area Multi-Object Fiber Spectroscopic Telescope) Galactic survey. A larger spectroscopical survey of the Galactic stars (Zhao et al. 2012) will provide higher probability to find more candidates as such. Meanwhile, future follow-up high-resolution spectroscopic observations of the 33 candidates are of great interest to confirm their high

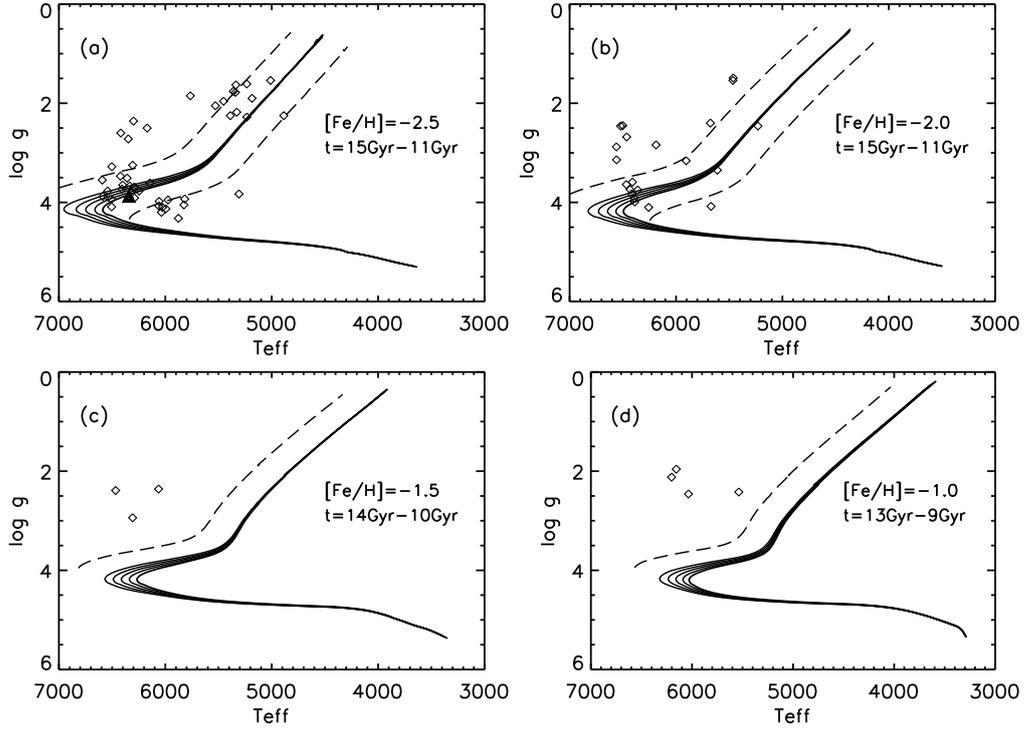

**Fig. 9** The log $g$ versus $T_{eff}$ diagram for four different metallicities [Fe/H] ($-1.0$, $-1.5$, $-2.0$ and $-2.5$) with a age coverage of 5 Gyrs:(a) [Fe/H] =$-3.0 \sim -2.25$, t=$11 \sim 15$ Gyrs; (b) [Fe/H] =$-2.25 \sim -1.75$, t=$11 \sim 15$ Gyrs; (c) [Fe/H] =$-1.75 \sim -1.25$, t=$11 \sim 15$ Gyrs; (d) [Fe/H] =$-1.25 \sim -1.0$, t=$11 \sim 15$ Gyrs. The isochrones corresponding to (a) $-$ (d) are shown in solid lines, and the dash line are the isochrones with a variation in $T_{eff}$ and log $g$ of 250 K and 0.5 $dex$, respectively.

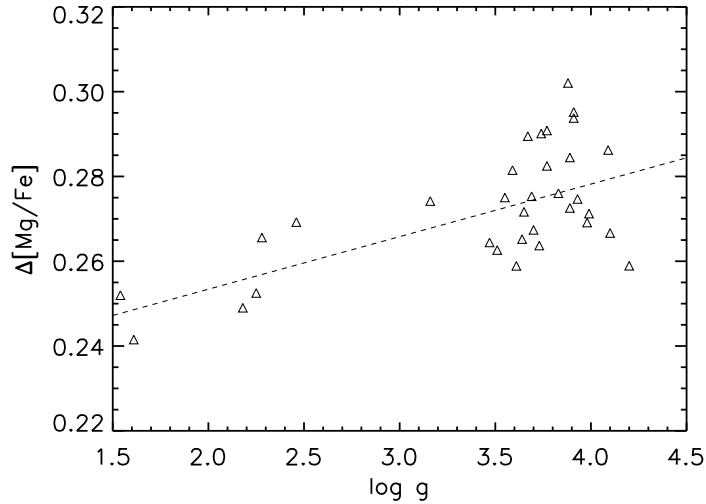

**Fig. 10** Variations of [Mg/Fe] stem from 0.5 $dex$ increase on log $g$, $\Delta$ [Mg/Fe]=[Mg/Fe]$_{logg}$-[Mg/Fe]$_{(logg+0.5)}$.

Table 3  Catalog of strongly Mg-enhanced candidates

| Star | Plate | MJD | Fiberid | $T_{eff}$ (K) | [Fe/H] | log g | RV (km/s) | S/N | [Mg/Fe] | Type |
|---|---|---|---|---|---|---|---|---|---|---|
| SDSS J001820.36-091833.0 | 652 | 52138 | 545 | 6334.59 | -2.61 | 3.67 | -57.71 | 51.33 | 1.11 | A |
| SDSS J012552.41+382358.4 | 2063 | 53359 | 130 | 6538.28 | -2.34 | 3.89 | -352.44 | 75.78 | 1.25* | A |
| SDSS J025432.96+354104.5 | 2378 | 53759 | 408 | 6400.75 | -2.53 | 3.65 | -274.01 | 56.12 | 1.18* | A |
| SDSS J084016.16+540526.4 | 2316 | 53757 | 515 | 6289.86 | -2.54 | 3.69 | -9.63 | 34.44 | 1.05 | A |
| SDSS J085650.28+401730.9 | 1199 | 52703 | 437 | 6246.56 | -2.34 | 3.77 | -28.42 | 47.48 | 1.17 | A |
| SDSS J094649.03+145432.5 | 2582 | 54139 | 407 | 6289.79 | -2.60 | 3.91 | 126.12 | 41.25 | 1.15 | A |
| SDSS J105018.63+000049.6 | 2389 | 54213 | 556 | 5233.95 | -2.67 | 2.28 | 312.59 | 38.39 | 1.10 | A |
| SDSS J110821.68+174746.5 | 2491 | 53855 | 389 | 6144.90 | -2.91 | 3.61 | -16.65 | 48.54 | 1.22* | A |
| SDSS J120231.14+204922.4 | 2893 | 54552 | 340 | 6388.34 | -2.07 | 3.99 | 147.55 | 47.46 | 1.06 | A |
| SDSS J125422.99+202619.5 | 2899 | 54568 | 332 | 6407.53 | -2.70 | 3.73 | 311.96 | 53.37 | 1.19 | A |
| SDSS J125712.60+592129.0 | 2446 | 54571 | 110 | 6393.55 | -2.16 | 3.91 | 35.83 | 52.11 | 1.15 | A |
| SDSS J130047.06+601828.3 | 2446 | 54571 | 626 | 6586.67 | -2.46 | 3.88 | -199.92 | 53.60 | 1.11 | A |
| SDSS J130538.01+194305.6 | 3235 | 54880 | 194 | 5230.05 | -2.25 | 2.46 | -100.78 | 54.67 | 1.00 | A |
| SDSS J134922.91+140736.9 | 1777 | 53857 | 479 | 6342.81 | -2.83 | 3.89 | -75.47 | 39.29 | 1.12 | A |
| SDSS J140038.27+230515.2 | 2784 | 54529 | 464 | 6431.97 | -2.07 | 3.74 | -39.70 | 50.77 | 1.12 | A |
| SDSS J140501.51+361759.9 | 2906 | 54577 | 307 | 6505.97 | -2.50 | 3.93 | -73.52 | 54.84 | 1.22* | A |
| SDSS J154120.53+085602.7 | 1724 | 53859 | 420 | 6036.23 | -2.51 | 4.20 | 25.75 | 40.86 | 1.02 | A |
| SDSS J164023.94+233349.5 | 1571 | 53174 | 617 | 6296.61 | -2.53 | 3.70 | -98.69 | 43.01 | 1.24* | A |
| SDSS J172813.66+081011.7 | 2797 | 54616 | 477 | 6417.26 | -2.16 | 3.83 | -187.82 | 46.85 | 1.07 | A |
| SDSS J233534.77+094331.5 | 2628 | 54326 | 380 | 6542.45 | -2.47 | 3.77 | 36.89 | 59.74 | 1.26* | A |
| SDSS J014419.25-084818.7 | 2816 | 54400 | 596 | 6592.59 | -2.30 | 3.55 | -106.03 | 67.45 | 1.16 | B |
| SDSS J015505.42-000421.1 | 2851 | 54485 | 356 | 5234.59 | -2.63 | 1.61 | -176.20 | 37.35 | 1.32* | B |
| SDSS J084444.69+063124.0 | 2317 | 54152 | 349 | 6359.19 | -2.57 | 3.51 | 224.77 | 46.30 | 1.17* | B |
| SDSS J120624.14+184411.2 | 2893 | 54552 | 129 | 6422.26 | -2.91 | 3.47 | -123.45 | 84.56 | 1.30* | B |
| SDSS J123850.76+173155.3 | 2599 | 54234 | 554 | 6466.74 | -2.19 | 3.64 | 17.74 | 48.53 | 1.04 | B |
| SDSS J131654.14+391830.4 | 3240 | 54883 | 351 | 6257.38 | -2.07 | 4.10 | 62.58 | 46.86 | 1.07 | B |
| SDSS J135718.30+194052.9 | 2770 | 54510 | 210 | 6504.55 | -2.53 | 4.09 | -77.95 | 54.28 | 1.15 | B |
| SDSS J161021.87+171130.1 | 2177 | 54557 | 382 | 6409.45 | -2.28 | 3.59 | -76.88 | 58.79 | 1.09 | B |
| SDSS J172229.03+270858.8 | 2182 | 53905 | 429 | 6058.25 | -2.52 | 3.98 | -79.94 | 52.19 | 1.09 | B |
| SDSS J172556.84+081101.8 | 2797 | 54616 | 383 | 5010.62 | -2.97 | 1.54 | -355.32 | 53.48 | 1.27* | B |
| SDSS J173113.88+334921.2 | 2253 | 54551 | 407 | 5328.50 | -2.85 | 2.18 | -228.16 | 50.61 | 1.25* | B |
| SDSS J204224.42-062424.7 | 1916 | 53269 | 243 | 5905.13 | -2.00 | 3.16 | 2.08 | 45.67 | 1.12 | B |
| SDSS J222617.34+010644.9 | 1144 | 53238 | 605 | 5388.60 | -2.43 | 2.25 | -258.87 | 55.46 | 1.08 | B |

Notes:  * Even under the extreme case of underestimation of log g by 0.5 $dex$, the [Mg/Fe] of these candidates are still larger than 1.0 $dex$.

Mg abundances, and detailed chemical abundances of other elements could help us to understand the origins of these peculiar stars and to further probe the formation and chemical evolution of the Galaxy.

**Acknowledgements** XL gives thanks to Lan Zhang, Jing Ren and Wei Wang for their helpful suggestions and to Andreas Koch for providing us the linearly interpolated program of atmospheric models, especial thanks also go to Poul Erik Nissen for kindly providing of FIES spectra. This work was supported by the National Natural Science Foundation of China under grant Nos. 11390371, 11233004, 11222326, and 11103030.

Funding for SDSS-III has been provided by the Alfred P. Sloan Foundation, the Participating Institutions, the National Science Foundation, and the U.S. Department of Energy Office of Science. The SDSS-III web site is http://www.sdss3.org/. SDSS-III is managed by the Astrophysical Research Consortium for the Participating Institutions of the SDSS-III Collaboration including the University of Arizona, the Brazilian Participation Group, Brookhaven National Laboratory, Carnegie Mellon University, University of Florida, the French Participation Group, the German Participation Group, Harvard University, the Instituto de Astrofisica de Canarias, the Michigan State/Notre Dame/JINA Participation Group, Johns Hopkins University, Lawrence Berkeley National Laboratory, Max Planck Institute for Astrophysics, Max Planck Institute for Extraterrestrial Physics, New Mexico State University, New York University, Ohio State University, Pennsylvania State University, University of Portsmouth, Princeton University, the Spanish Participation Group, University of Tokyo, University of Utah, Vanderbilt University, University of Virginia, University of Washington, and Yale University.